# Analysis and design of multilayer structures for neutron monochromators and supermirrors

S. Masalovich

Technische Universität München, Forschungsneutronenquelle Heinz Maier-Leibnitz (FRM II), Lichtenbergstr. 1, D-85747 Garching, Germany
E-mail: Sergey.Masalovich@frm2.tum.de

**Abstract**

A relatively simple and accurate analytical model for studying the reflectivity of neutron multilayer monochromators and supermirrors is proposed. Design conditions that must be fulfilled in order to reach the maximum reflectivity are considered. The question of the narrowest bandwidth of a monochromator is discussed and the number of layers required to build such a monochromator is derived. Finally, we propose a new and efficient algorithm for synthesis of a supermirror with specified parameters and discuss some inherent restrictions on an attainable reflectivity.

# 1. Introduction

Multilayer structures have found a wide application in neutron instrumentation as monochromators, polarizes and supermirrors [1, 2]. The latter ones, for instance, are commonly used nowadays at research reactors for construction of neutron guides with enhanced angular acceptance designed to transport neutrons over long distances. Multilayer supermirrors find use also in neutron focusing devices [3 - 5] opening a new way in neutron instrumentation.

Generally, a multilayer structure represents a thin film system composed of layers of two different materials alternatively and repeatedly deposited on a flat substrate (Fig. 1).

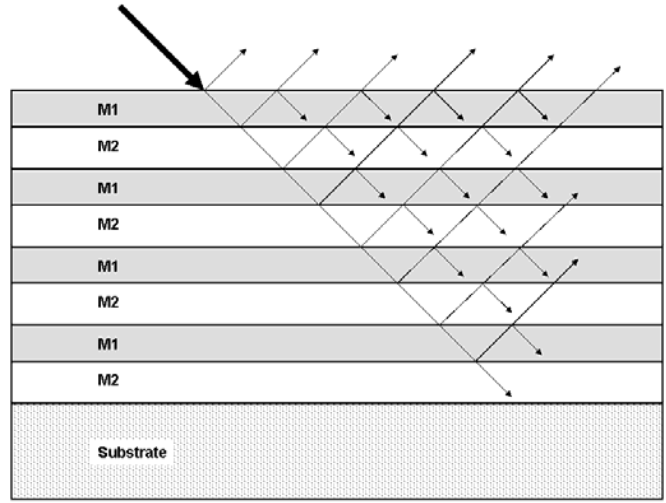

Fig. 1. View of the multilayer structure composed of layers of two different materials M1 and M2 deposited on a flat substrate. The multiple waves are shown schematically (see text).

These materials are chosen to have high and low effective potentials for neutrons (also known as neutron optical potentials) and, therefore, a multilayer system can be considered as a sequence of one-dimensional square-well potentials. During propagation through the system a partial reflection and transmission of a neutron wave occurs at every interface resulting in appearance of multiple waves within the system (see Fig. 1). In the case of a periodic structure, the multiple wave interference leads to distinctive band structures in the energy spectrum of the beams reflected from or transmitted through the multilayer. Hence, with a proper selection of the layer materials and thicknesses one can, in principle, built a system with desired spectral properties. In the present paper we propose a relatively simple and accurate analytical method for studying the reflectivity, $R$, of multilayer mirrors and apply this method for design of two main systems which have an extensive application in neutron instrumentation: (a) a very narrow bandwidth system for a neutron monochromator and (b) a very wide bandwidth system for a neutron supermirror. We begin with the multilayer structure made up as a sequence of identical bilayers repeated many times in one direction where each bilayer consists of two thin films of different materials. Such systems are widely used as monochromators in neutron optics (see, e.g., [6 - 9]). We shall find conditions that must be fulfilled in order to attain the maximum reflectivity for neutrons with a fixed incident wave vector. We evaluate then the bandwidth and the number of bilayers required to build such a monochromator.

We next apply obtained results for the synthesis of a neutron supermirror. In that system successive bilayers vary gradually in thickness in such a way that the neutron reflectivity displays a very wide bandwidth [10 -19]. We introduce a new and efficient algorithm for design of a supermirror with specified parameters and discuss some inherent limitations on an attainable reflectivity.

## 2. General remarks

For a given material an effective potential, $U$, is defined as

$$U = \frac{2\pi\hbar^2}{m_n} \cdot \sum_j N_j b_j, \qquad (1)$$

where $m_n$ is the neutron mass, $b_j$ is the bound coherent scattering length and $N_j$ is the number density of nuclei. The summation runs over all elements and isotopes that constitute the layer. It is worth noting that in magnetic materials the effective potential will also include a magnetic interaction of neutrons with matter. This obviously opens a way for construction of polarizing devices. In the present paper we do not give a special consideration to that case since all formulae for polarizing systems can be obtained straightforward from our results derived for the general case of wave propagation in a one-dimensional potential.

It is evident that neutron waves propagating through a multilayer structure undergo multiple reflections at interfaces and the resulting reflectivity and transmittance of the system are determined by the interference of all the multiple reflected waves. The interference pattern depends apparently on the phases of summed waves and, thus, on the thicknesses of the layers and the magnitudes of the neutron wave vectors within the layers. The latter ones are defined by the following expression:

$$k = \sqrt{\frac{2m_n}{\hbar^2}(E_0 - U)}, \qquad (2)$$

where $E_0$ is the energy of an incident neutron in vacuum. Generally, in order to calculate the phase of the wave at a given point one needs to know the magnitude of the wave vector and the direction of the wave propagation. However, for a one-dimensional potential structure the problem can be simplified significantly. Indeed, in that case the components of the neutron momentum which are parallel to the multilayer surface do not vary when the neutron travels through the interface between two different media. Thus, these components of the neutron wave vector and the part of the neutron energy associated with those components are constant and they can be omitted from the subsequent consideration. As a result the wave propagation through the system can be characterized merely by the component normal to the multilayer surface (see, e.g., [13, 17, 20])

$$k_\perp = \sqrt{\frac{2m_n}{\hbar^2}(E_{0\perp} - U)} \qquad (3)$$

with $E_{0\perp}$ being the part of the incident neutron energy that corresponds to this component. Therefore, we reduce the problem of finding the specular reflection coefficient of neutrons with the wave vector $\vec{k} = (\vec{k}_\parallel, \vec{k}_\perp)$ incident upon a multilayer to the problem of finding the specular reflection coefficient of neutrons with the incident wave vector $\vec{k} = \vec{k}_\perp$. For simplicity we shall omit the sub index "⊥" from our subsequent calculations keeping in mind that only components normal to the surfaces (interfaces) are considered.

## 3. Multilayer system composed of identical bilayers: neutron monochromator

We assume first that the multilayer system is composed of thin films of non-absorbing and non-scattering materials. This is a reasonably good approximation since in most cases absorption and scattering are very low and can be initially neglected. We shall discuss their effect on the reflectivity later when we present our results obtained for supermirrors. Next, we postulate that in the case of the total reflection (i.e., $R=1$) the neutron flux through any plane, which is located within the multilayer parallel to the surface, has to be equal zero. This postulate looks obvious and we apply it below to study the reflectivity from a multilayer system.

To find the flux within a multilayer we have to solve the quantum-mechanical problem of a neutron wave traveling though the system with one-dimensional periodical potential (see Fig. 2).

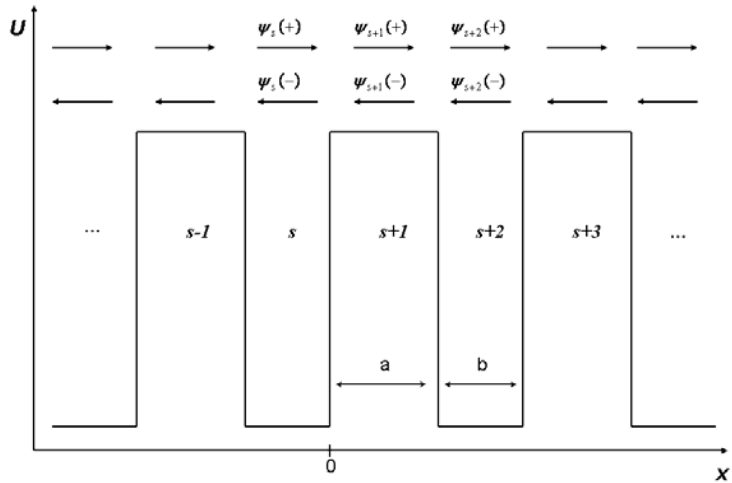

Fig. 2. Effective potential $U$ as a function of the distance $x$ along the direction normal to the multilayer surface. Arrows at the top indicate the waves propagating in the layers. $a$ and $b$ – thicknesses of the layers.

In quantum mechanics a flux, $F$, is defined as

$$F = \frac{i\hbar}{2m}\left(\psi \cdot \nabla \psi^* - \psi^* \cdot \nabla \psi\right), \tag{4}$$

where $\psi$ and $\nabla\psi$ are the neutron wave function and its gradient and the asterisk denotes complex conjunction. In the particular case of the one-dimensional potential the gradient becomes merely a derivative along the normal to the interface. The wave function within the thickness of any layer can be written in a common way (see Fig. 2)

$$\psi = \psi(+) + \psi(-) \equiv A\exp(ikx) + B\exp(-ikx) \tag{5}$$

Here $A$ is the amplitude of the wave $\psi(+)$ traveling in the direction of the incident neutron wave and $B$ – the amplitude of the wave $\psi(-)$ traveling in the opposite direction; $k$ is the magnitude of the wave vector within the layer. On substituting Eq. (5) into Eq. (4), we obtain

$$F = \frac{kh}{m}\left(|A|^2 - |B|^2\right). \tag{6}$$

We define now a reflectance amplitude, $r$ (a priori complex), within a layer as

$$|r| = \left|\frac{B}{A}\right| \tag{7}$$

From Eqs. (6) and (7) it follows that the condition $F = 0$ holds only if $|r|^2 = 1$. Thus, using the postulate mentioned above one may conclude that $|r|^2 = 1$ is required in order that $R = 1$. The evaluation of the parameters of the multilayer system that ensure $|r|^2 = 1$ constitutes the main subject of our subsequent calculations.

First we discuss the reflection of the neutron wave from a semi-infinite multilayer in which the number of layers is infinite in one direction. Let us consider three subsequent layers within the multilayer (Fig. 2) which have numbers $s$, $s+1$ and $s+2$. By analogy with Eq. (5) one can write the neutron wavefunction in each layer as a sum of the waves traveling to the right and to the left:

$$\psi_s = A_s \exp(ik_s x) + B_s \exp(-ik_s x) \tag{8a}$$

$$\psi_{s+1} = A_{s+1} \exp(ik_{s+1} x) + B_{s+1} \exp(-ik_{s+1} x) \tag{8b}$$

$$\psi_{s+2} = A_{s+2} \exp(ik_{s+2} x) + B_{s+2} \exp(-ik_{s+2} x) \tag{8c}$$

Here a subscript index was used to identify a layer. We chose the system of coordinates with $x = 0$ at the interface between the layers $s$ and $s + 1$. The requirement of the continuity of the wave function and its derivative at the interfaces $x = 0$ and $x = a$ (see Fig. 2) leads to four equations with six unknown parameters $A_{s,s+1,s+2}$ and $B_{s,s+1,s+2}$.

$$A_s + B_s = A_{s+1} + B_{s+1} \tag{9a}$$

$$k_s A_s - k_s B_s = k_{s+1} A_{s+1} - k_{s+2} B_{s+2} \tag{9b}$$

$$A_{s+1}\exp(ik_{s+1}a)+B_{s+1}\exp(-ik_{s+1}a)=A_{s+2}\exp(ik_{s+2}a)+B_{s+2}\exp(-ik_{s+2}a) \qquad (9c)$$

$$k_{s+1}A_{s+1}\exp(ik_{s+1}a)-k_{s+1}B_{s+1}\exp(-ik_{s+1}a)=k_{s+2}A_{s+2}\exp(ik_{s+2}a)-k_{s+2}B_{s+2}\exp(-ik_{s+2}a) \qquad (9d)$$

where *a* is the thicknesses of the *(s+1)* layer.

We are to solve the system (9) for the parameter $r = B_s/A_s$. Therefore, without loss of generality one may divide all equations in (9) by $A_s$ reducing the number of unknown parameters to five. Now, to be able to solve the system one more equation is needed. We shall try to derive this equation using the fact that the multilayer system represents a periodic structure infinite in one direction. From this periodicity we may assume that the same ratio of the amplitudes of the waves traveling to the left and to the right will take place at identical interfaces. This assumption plays a central role in our calculations and its justification lies in the successful design of multilayer systems verified by direct matrix calculations. We apply this assumption to the waves at the interfaces $x = 0$ and $x = a + b$ (Fig. 2). At $x = 0$ in the layer "s" we have $r = B_s/A_s$. Now, the amplitude of the wave incident upon the interface $x = a + b$ is $A_{s+2}\exp(ik_{s+2}(a+b))$ and the amplitude of the reflected wave is $B_{s+2}\exp(-ik_{s+2}(a+b))$. Then, according to the assumption one can write

$$\frac{B_{s+2}}{A_{s+2}}\exp(-i\cdot 2k_{s+2}(a+b))=\frac{B_s}{A_s} \qquad (10)$$

The set of equations (9) combined with Eq. (10) can now be solved for the parameters $r = B_s/A_s$ and $C = A_{s+2}/A_s$ which are of primary interest to us. Analysis of the first parameter, $|r|^2$, allows one to draw the conditions under which the total reflection from the multilayer occurs. The second parameter $|C|^2 \equiv |A_{s+2}/A_s|^2$ shows the variation of the neutron intensity from bilayer to bilayer in the direction of an incident wave and can be used for further analysis of the system. We begin with the analysis of the reflectivity.

### 3.1. Reflectivity of a multilayer monochromator

With the aim to generalize our calculations we introduce here a parameter *m* defined as

$$m = \frac{k_0}{k_c}. \qquad (11)$$

Here $k_0$ is the normal component of an incident wave vector in vacuum and $k_c$ is the critical wave vector associated with a given material. The last quantity can be understood if we consider the condition of total specular reflection from the ideal flat surface of a bulk medium. Such reflection may occur only if the part of the neutron energy associated with the movement normal to the surface is less than the potential barrier of the medium (see Eq. (2)). The $k_c$ can be then derived directly from the condition $E_0 = \hbar^2 k_c^2/2m_n = U$. Thus, we obtain

$$k_c^2 = 4\pi \cdot \sum_j N_j b_j \equiv 4\pi \cdot (Nb), \tag{12}$$

where we used $(Nb) = \sum_j N_j b_j$ for simplicity. The condition of total specular reflection from a bulk medium can be then written as

$$k_0 < k_c \text{ or } m < 1. \tag{13}$$

The situation changes drastically if we consider the reflection from a thin-film multilayer structure. In this case the interference between the waves multiple reflected from different interfaces leads to specific features in reflectivity and, in principal, Eq. (13) does not hold. In particular, the interference gives rise to specular reflection of neutrons with $k_0 \geq k_c$ where $k_c$ is defined by the material with the highest potential (layers (s+1), (s+3) and so on in Fig. 2). Below we study the reflectivity of a multilayer by solving the set of equations (9-10).

One can show that the solution to the set (9-10) for $r$ can be found as roots of the following equation:

$$y^2 + Q \cdot y + 1 = 0. \tag{14}$$

Here

$$y = r \cdot \exp(ik_b b) \tag{15}$$

and

$$Q = \frac{(k_a - k_b)^2}{(k_a^2 - k_b^2)} \cdot \frac{\sin(k_a a - k_b b)}{\sin(k_a a)} + \frac{(k_a + k_b)^2}{(k_a^2 - k_b^2)} \cdot \frac{\sin(k_a a + k_b b)}{\sin(k_a a)} =$$
$$= \frac{4 k_a k_b}{(k_a^2 - k_b^2)\sin(k_a a)} \left( \frac{(k_a^2 + k_b^2)}{2 k_a k_b} \sin k_a a \cdot \cos k_b b + \cos k_a a \cdot \sin k_b b \right) \tag{16}$$

For notational convenience we used here the substitutions: $k_{s+1} \equiv k_a$ and $k_{s+2} \equiv k_b$. Notice, that the quantities $k_a$ and $k_b$ for a given multilayer and for fixed $k_0$ can be evaluated with the use of the parameters $k_c$ and $m$ as follows

$$k_a^2 = k_0^2 - 4\pi(Nb)_a = 4\pi(Nb)_a \cdot (m^2 - 1) = k_c^2(m^2 - 1) \tag{17}$$

$$k_b^2 = k_0^2 - 4\pi(Nb)_b = 4\pi(Nb)_b \cdot (m^2 - \rho) = k_c^2(m^2 - \rho) \tag{18}$$

with

$$\rho = \frac{(Nb)_b}{(Nb)_a}. \tag{19}$$

On writing the solution to Eq. (14) in the form

$$y = -\frac{Q}{2} \pm \sqrt{\frac{Q^2}{4} - 1}, \tag{20}$$

we may conclude that $|y|$ becomes unity and consequently $R = |r|^2 = |y|^2 = 1$ if and only if the following condition

$$Q^2 \leq 4 \tag{21}$$

is fulfilled.
Substitution Eq. (16) for $Q$ in Eq. (21) gives

$$\left(\frac{k_a - k_b}{k_a + k_b}\right)^4 \sin^2(k_a a - k_b b) + \sin^2(k_a a + k_b b) \leq 2 \cdot \left(\frac{k_a - k_b}{k_a + k_b}\right)^2 \left(\sin^2 k_a a + \sin^2 k_b b\right). \tag{22}$$

Thus, to achieve total reflection for a fixed incident wave vector one has to find a pair of thicknesses $a$ and $b$ that satisfy (22). Fig. 3 illustrates the numerical solution (black areas) to the inequality (22) in the parameter space $(a,b)$ for a Ni/Ti multilayer system at $m=3$.

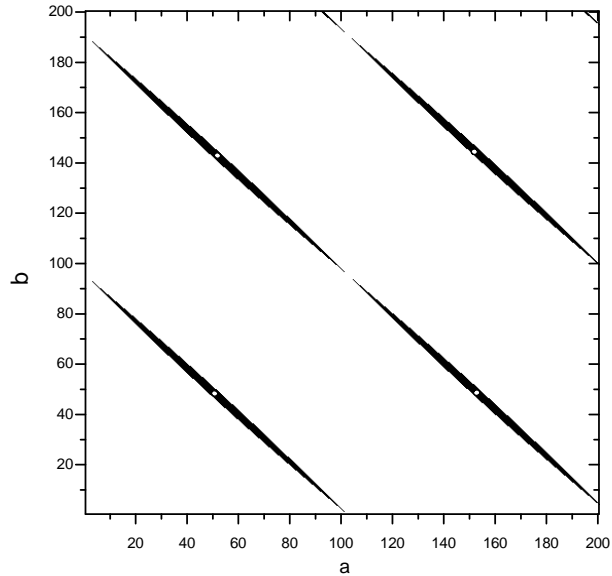

Fig. 3. Graphical representation of the solution to the inequality (22) in the $(a,b)$ plane for a Ni/Ti multilayer and for incident neutrons with $m=3$. The condition $R=1$ holds at all points in the black areas. The thicknesses of Ni-layer ($a$) and Ti-layer ($b$) are given in Å.

Looking at Fig. 3, we see that all points *(a,b)* where the reflectivity is equal to one constitute continuous areas arranged regularly in the *(a,b)* plane. This regular pattern obviously results from the fact that the quantity Q contains periodic functions. The middle line of each area (not shown) corresponds to $Q = 0$, which can be written as

$$\left(k_a^2 + k_b^2\right) \cdot \sin k_a a \cdot \cos k_b b + 2 k_a k_b \cdot \cos k_a a \cdot \sin k_b b = 0 \quad (k_a a \neq n\pi,\ n\text{-integer}). \tag{23}$$

## 3.2. Quarter-wave layers

There are unique points in Fig. 3 (marked as white points in the middle of each black area) where

$$k_a a = \frac{\pi}{2} + n\pi, \quad k_b b = \frac{\pi}{2} + n\pi \quad (n\text{-integer})$$

At these points the condition $Q = 0$ is obviously fulfilled (see Eqs. (16) and (23)) and thus $Q^2 < 4$ holds. Multilayer systems composed of bilayers with $k_a a = \pi/2$ and $k_b b = \pi/2$ are known in optics as quarter-wave mirrors and quarter-wave filters. They are broadly used in neutron instrumentation as well for production of neutron monochromators and supermirrors. The advantage of using quarter-wave layers is clearly seen in Fig. 3. Indeed, around the point where $k_a a = \pi/2$ and $k_b b = \pi/2$ the width of the area with $R = 1$ in the *(a,b)* plane is maximal. Hence, a small systematical error in the layer thickness (*a* or/and *b*) during production process will not cause pronounced degradation of the reflectivity at a given *m*. Since it is important for both design and production work, we consider here this statement in more details.

    Let us take an arbitrary value *a* (Ni-layer thickness) from the interval related to the first $R = 1$ area in the *(a,b)* plane computed for fixed *m*. We chose this interval for the practical reason that the amount of material necessary for production of a multilayer structure is minimal. So, for this fixed *a* the condition (23) allows us to find the parameter $b_a = b(a)$ such that the point $(a, b_a)$ lies on the middle line of the area with $R = 1$. One can evaluate now the interval $[b_a - \Delta b_a,\ b_a + \Delta b_a]$ where the thickness of the Ti-layer meets the condition $R = 1$. The boundaries of this interval can be found by solving the equation $Q^2(a, b_a \pm \Delta b_a) = 4$ for $\Delta b_a$. We omit here direct but lengthy calculations and write the solution in the form

$$k_b |\Delta b_a| = \frac{(1-\rho) \cdot \sin k_a a}{\sqrt{4(m^2 - 1)(m^2 - \rho) + (1-\rho)^2 \sin^2 k_a a}}. \tag{24}$$

It is easy to verify that $|\Delta b_a|$ reaches its maximum when $\sin k_a a = 1$, that is when the parameter *a* satisfies the condition $k_a a = \pi/2$. Since at this point $k_b b_a = \pi/2$ holds as well, we can modify Eq. (24) into

$$\frac{|\Delta b_a|}{b_a} = \frac{2}{\pi} \cdot \frac{(1-\rho)}{\sqrt{4(m^2-1)(m^2-\rho)+(1-\rho)^2}} = \frac{2}{\pi} \cdot \frac{(1-\rho)}{(2m^2 - 1 - \rho)} \quad \text{(quarter-wave layers)}. \tag{25}$$

Eq. (25) shows the upper limit of the acceptable systematical error in the layer thickness when producing a multilayer with $R=1$ for a specified $m$. For the sake of illustration, in Table 1 the quarter-wave Ti-layer thickness ($b$), relative ($\Delta b/b$) (Eq. (25)) and absolute value ($\Delta b$) are computed for different $m$.

Table 1
Comparison of the quarter-wave Ti-layer parameters for different $m$.

| $m$ | $b$ [Å] | $\Delta b/b$ | $\Delta b$ [Å] |
| --- | --- | --- | --- |
| 1.5 | 92.16 | 0.207 | 19.10 |
| 2.0 | 70.42 | 0.107 | 7.51 |
| 2.5 | 56.85 | 0.066 | 3.73 |
| 3.0 | 47.61 | 0.045 | 2.13 |
| 3.5 | 40.92 | 0.032 | 1.33 |
| 4.0 | 35.88 | 0.025 | 0.88 |
| 4.5 | 31.94 | 0.019 | 0.62 |
| 5.0 | 28.77 | 0.016 | 0.45 |

It is worth noting that for $m \approx 2.5$ and greater the acceptable deviation $\Delta b$ presented in Table 1 is less than the Ti lattice constant ($a_{Ti} = 2.95\text{Å}, c_{Ti} = 4.68\text{Å}$). Therefore, we may expect that, from this point on, the total specular reflection for a specified $m$ becomes unattainable. However, for low-$m$ monochromators the acceptable deviation $\Delta b$ seems to be relatively large and allows the highest reflectivity to be reached.[1]

In the present work we do not carry out special calculations to obtain the relation like Eq. (25) for the Ni-layers. Nevertheless, from Fig. 3 it becomes clear that the acceptable error in the layer thickness $\Delta a$ in that case is very similar to what was obtained for the Ti-layers.

### 3.3. Wave penetration into a multilayer

Let us consider now the parameter $|C|^2 \equiv |A_{s+2}/A_s|^2$ which defines the variation of the neutron intensity as it propagates through the multilayer. From Eq. (9) we can derive

$$i \cdot r \cdot (k_a^2 - k_b^2) \cdot \sin k_a a + i \cdot (k_a^2 + k_b^2) \cdot \sin k_a a + 2k_a k_b \cdot \cos k_a a = \\ = 2k_a k_b \cdot C \cdot \exp(ik_b a) \quad (26)$$

which will be used for further evaluations. First we analyze this equation for those parameters $a$ and $b$ that satisfy $Q^2 > 4$.

---

[1] The model under consideration does not include interfacial roughness, interdiffusion and the formation of intermetallic compounds at the Ni-Ti interfaces which generally result in degradation of the multilayer performance.

### 3.3.1. Case $Q^2 > 4$

Looking at Eq. (20), we conclude that in this case *y* becomes a real number. Thus, one can write

$$r = y \cdot \exp(-ik_b b) = y \cdot \cos k_b b - i \cdot y \cdot \sin(k_b b). \tag{27}$$

Substitution of Eq. (27) into (26) gives:

$$|C|^2 = 1 + (y^2 + 1)\frac{(k_a^2 - k_b^2)^2}{4k_a^2 k_b^2} \sin^2 k_a a + \\ + y\frac{k_a^2 - k_b^2}{k_a k_b} \sin k_a a \cdot \left(\frac{k_a^2 + k_b^2}{2k_a^2 k_b^2} \sin k_a a \cdot \cos k_b b + \cos k_a a \cdot \sin k_b b\right). \tag{28}$$

Now, from Eqs. (14) and (16) we obtain

$$y^2 + 1 = -Q \cdot y \tag{29}$$

$$\left(\frac{k_a^2 + k_b^2}{2k_a k_b} \sin k_a a \cdot \cos k_b b + \cos k_a a \cdot \sin k_b b\right) = Q\frac{k_a^2 - k_b^2}{4k_a k_b} \sin k_a a \tag{30}$$

Finally, the substitution of Eqs. (29) and (30) into Eq. (28) results in

$$|C|^2 = 1 \quad (Q^2 > 4)$$

We see that the part of the neutron intensity that penetrates into a multilayer system passes through unaffected (obviously, it is true only for ideal materials free from absorption and scattering).

### 3.3.2. Case $Q^2 \leq 4$

We now turn our attention to the case $Q^2 \leq 4$, that is, to the case of the total reflection. As mentioned above, the condition $|r| = 1$ holds and therefore we can write

$$r = \exp(i\varphi) = \cos\varphi + i\sin\varphi \tag{31}$$

where φ is an arbitrary phase.
Combining Eqs. (31) and (26) yields

$$|C|^2 = 1 + \frac{(k_a^2 - k_b^2)^2}{2k_a^2 k_b^2} \sin^2 k_a a +$$
$$+ \frac{k_a^2 - k_b^2}{k_a k_b} \sin k_a a \cdot \left( \frac{k_a^2 + k_b^2}{2k_a k_b} \sin k_a a \cdot \cos\varphi - \cos k_a a \cdot \sin\varphi \right) \quad (32)$$

Notice, that in the case of total reflection no part of the neutron intensity propagates through the structure at infinity. It means that the neutron wave function falls off within the multilayer and one may expect that $|C|^2 < 1$. We are interested to find such parameters of the thin-film structure which would provide the smallest $|C|^2$ and, hence, the most rapid attenuation of the wave function. Obviously, in that case the number of layers required to ensure the total reflection would be minimal. For the same practical reason that was already mention earlier we consider only the points $(a,b)$ that belong to the first area with $R=1$ in the $(a,b)$ plane (see, e.g., Fig. 3). At all these points the condition $\sin k_a a > 0$ holds. Then, from Eq. (32) we see that for fixed $k_a a$ there exists $\varphi$ that minimizes $|C|^2$. Since in our consideration $k_a < k_b$, one may conclude that $|C|^2$ reaches its minimum at the point where the function

$$f(\varphi) = \left( \frac{k_a^2 + k_b^2}{2k_a k_b} \sin k_a a \cdot \cos\varphi - \cos k_a a \cdot \sin\varphi \right) \quad (33)$$

reaches its positive maximum. It is easy to show that it happens if

$$\tan\varphi = -\frac{2k_a k_b}{k_a^2 + k_b^2} \cot k_a a \quad (34)$$

with the additional condition that $\cos\varphi > 0$.

Consequently, Eq. (32) reduces to

$$|C|^2 = 1 - \frac{(k_b^2 - k_a^2)^2}{2k_a^2 k_b^2} \sin k_a a \cdot \left[ \sqrt{\sin^2 k_a a + \frac{4k_a^2 k_b^2}{(k_b^2 - k_a^2)^2}} - \sin k_a a \right] \quad (35)$$

It follows from Eq. (35) that $|C|^2$ reaches its minimum at the point where $\sin k_a a = 1$ and hence $k_a a = \pi/2$. The last condition defines the parameter $a$. Now, we are to find the second parameter, $b$, which corresponds to the minimum of $|C|^2$. This can be done by substituting $k_a a = \pi/2$ into Eq. (34). We obtain then $\tan\varphi = 0$ and, consequently, $r = \exp(i\varphi) = 1$. Thus, Eq. (15) becomes

$$y = \exp(ik_b b) = \cos k_b b + i \sin k_b b \quad (36)$$

When this result is compared with Eq. (20) written in the form

$$y = -\frac{1}{2}Q \pm i\sqrt{1 - \frac{Q^2}{4}}, \qquad (37)$$

the following relationship is obtained: $\cos k_b b = \pm Q/2$.
On the other hand, the substitution $k_a a = \pi/2$ into Eq. (30) gives

$$\frac{Q}{2} = \frac{k_a^2 + k_b^2}{k_a^2 - k_b^2} \cdot \cos k_b b. \qquad (38)$$

So we may conclude that Eq. (38) holds only if $Q = 0$ and, consequently, $\cos k_b b = 0$ or $k_b b = \pi/2$. The last condition defines the parameter $b$. Thus, we have shown that for fixed $k_0$ the function $|C|^2$ reaches its minimum at the point where $k_a a = \pi/2$ and $k_b b = \pi/2$. To summarize, the multilayer structure built of quarter-wave layers features the least wave penetration into the system. The minimal value of $|C|^2$ in that case can be evalueted by substituting $k_a a = \pi/2$ into Eq. (35). We obtain then

$$|C|^2_{min} = \frac{k_a^2}{k_b^2} \qquad \text{(quarter-wave layers)} \qquad (39)$$

Taking into account Eqs. (17-18), we may write as well

$$|C|^2_{min} = \frac{m^2 - 1}{m^2 - \rho} \qquad \text{(quarter-wave layers)} \qquad (40)$$

This is the central result of our calculations and it will be applied to optimal design of various multilayer structures discussed in the present paper. First, from Eqs. (39-40) we immediately conclude that the smallest $|C|^2$ can be attained for a pair of materials where one material has negative (Nb) while the other has large positive (Nb). The very well known pair of that kind is nickel and titanium with $(Nb)_{Ni} = 9.41 \cdot 10^{-6}$ Å$^{-2}$ and $(Nb)_{Ti} = -1.95 \cdot 10^{-6}$ Å$^{-2}$. For this pair $\rho = -0.207$ and in the case of $m=3$ we obtain $|C|^2_{min} = 0.869$. We emphasize that this value characterizes an ideal system composed of quarter-wave layers. If, for instance, the thickness of one of the layers differs from the quarter-wave thickness then $|C|^2$ increases as shown illustratively in Fig. 4. In this figure $|C|^2$ is plotted as a function of the Ni-layer thickness with the Ti-layer being the true quarter-wave layer.

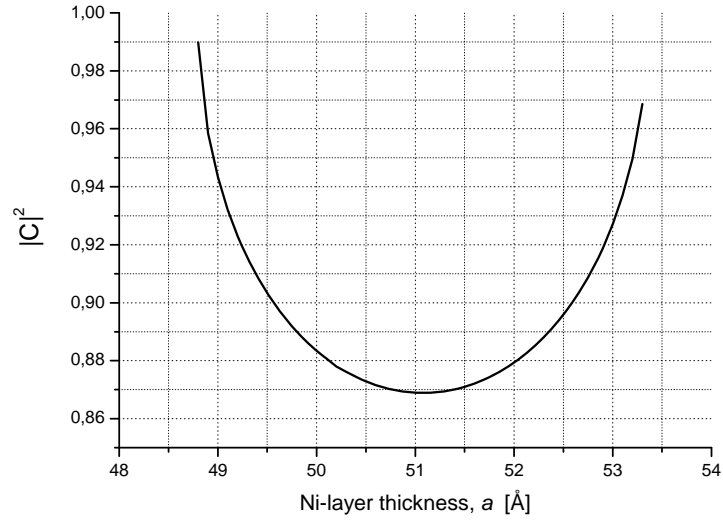

Fig. 4. $|C|^2$ as a function of the Ni-layer thickness $a$ calculated for the multilayer with $m = 3$. The Ti-layer thickness $b$ is fixed at $b = 47.605$Å to meet the quarter-wave condition.

As expected, $|C|^2$ reaches its minimum value 0.869 at the point $a = 51.071$Å where the Ni-layer comprises a quarter-wave layer.

It should be remembered that these results have been obtained for the multilayer structure with an infinite number of layers. With the aim to check the applicability of our model to a realistic multilayer system with a limited number of layers we performed direct calculation of the wave function distribution within such a system. One example is shown in Fig. 5 where the distribution of the squared modulus of the neutron wavefunction over the first layers of a Ni/Ti quarter-wave multilayer is presented. The multilayer consists of 90 bilayers deposited on a Si substrate. The result has been obtained with the use of the general calculation technique known as matrix method.

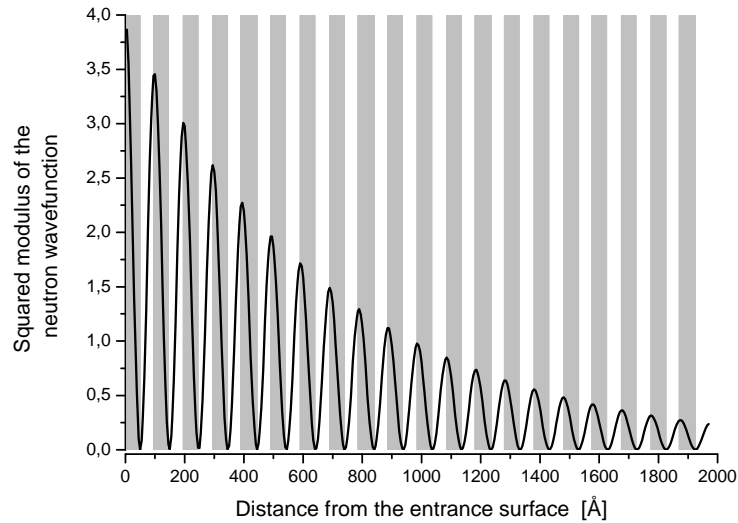

Fig. 5. Distribution of the squared modulus of the neutron wavefunction within the Ni/Ti quarter-wave multilayer with 90 bilayers designed for *m=3*. The Ni and Ti layers are shown as the gray and white stripes respectively.

The ratio of the amplitudes of the subsequent peaks in Fig. 5 checks well with the value 0.869 obtained earlier. The good agreement between the matrix calculations and the result obtained analytically for the semi-infinite multilayer confirms the validity of our theoretical model.

**3.4. Bandwidth of a multilayer monochromator**

Next we investigate the reflectivity as a function of the incident neutron wave vector. As mentioned at the beginning of this paper, only the component of the neutron wave vector normal to the multilayer surface will be considered. As an example, Fig. 6 shows the reflectivity spectrum calculated for the Ni/Ti quarter-wave multilayer studied in the previous section. The calculation has been done with the use of the matrix method.

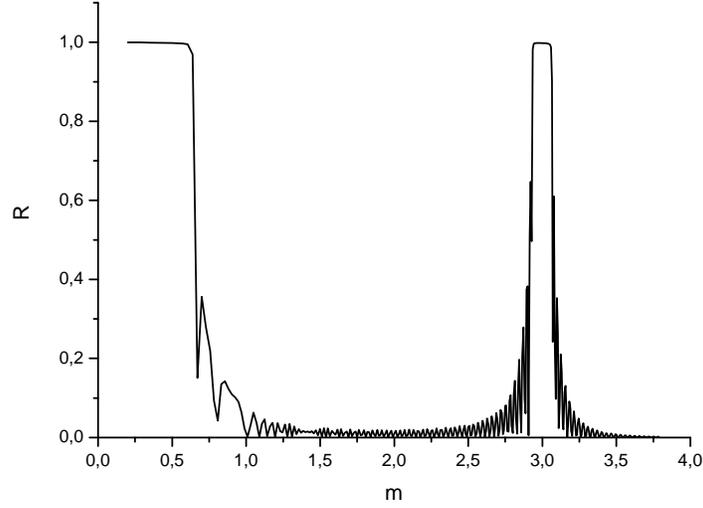

Fig. 6. Reflectivity of the Ni/Ti quarter-wave multilayer as a function of *m*. The multilayer consists of 90 bilayers on a Si substrate.

The fine structure of the reflectivity curve arises from the multiple wave interference. Usually, this fine structure is smoothed off by a spectral resolution function of a detector and thus not observable in an experiment. For us it is important to point out that the reflectivity curve in Fig. 6 reaches unity not only at the preset point *m=3*, but in some region around that point where it makes up a plateau. This plateau is analogous to the well known Darwin's plateau which appears in diffraction by single crystals. With the aim to make an estimate of the width of that plateau we turn again to the conditions (21 - 22). They say that the reflectivity of a semi-infinite multilayer is equal to one for any set of parameters $k_a$, $k_b$, *a* and *b* that satisfy the inequality (22). We have already seen that for fixed $k_a$ and $k_b$ (related to some fixed $k_0$) there are continuous areas in the *(a,b)* plane where (22) holds. By analogy, from the look of (22) one can assume that continuous areas exist as well in the ($k_a$, $k_b$) plane for fixed *a* and *b*. Taking into account Eqs. (17-18), we conclude that such areas would manifest themselves as plateaus on reflectivity curves when plotted against *m*. This is clearly seen in Fig. 6. We evaluate now the width of the plateau.

Let us consider a quarter-wave multilayer designed for $m = m_0$. Thus, we have $k_a(m_0) \cdot a = \pi/2$ and $k_b(m_0) \cdot b = \pi/2$. We now modify Eq. (16) as follows

$$Q(m) = \frac{g(m)}{h(m)}, \tag{41}$$

where

$$g(m) = (k_a(m) - k_b(m))^2 \cdot \sin(k_a(m)a - k_b(m)b) + (k_a(m) + k_b(m))^2 \cdot \sin(k_a(m)a + k_b(m)b) \tag{42}$$

$$h(m) = (k_a^2(m) - k_b^2(m)) \cdot \sin(k_a(m)a) \tag{43}$$

and

$$g(m_0) = 0 \tag{44}$$

$$h(m_0) = \left(k_a^2(m_0) - k_b^2(m_0)\right) \tag{45}$$

From the Taylor series expansion of $Q(m)$ around $m_0$ one can obtain

$$Q(m) = Q(m_0 + \Delta m) = Q(m_0) + Q'(m_0)\Delta m = Q'(m_0)\Delta m \tag{46}$$

where we keep only the linear terms. In Eq. (46) the first derivative of the function $Q(m)$ at the point $m = m_0$ can be written as

$$Q'(m_0) = \frac{g'(m_0)}{h(m_0)} \tag{47}$$

with

$$g'(m_0) = (k_a(m_0) - k_b(m_0))^2 (k'_a(m_0)a - k'_b(m_0)b) - (k_a(m_0) + k_b(m_0))^2 (k'_a(m_0)a + k'_b(m_0)b) \tag{48}$$

Next, we rearrange Eq. (21)

$$(g'(m_0) \cdot \Delta m)^2 \leq 4h^2(m_0) \tag{49}$$

and finally obtain the half-width of the plateau ($\Delta m$)

$$\Delta m = 2\left|\frac{h(m_0)}{g'(m_0)}\right| \tag{50}$$

In the case of a quarter-wave multilayer we have

$$g'(m_0) = \frac{\pi}{2} \cdot \frac{m \cdot k_c^2}{k_a^2 k_b^2} \left[(k_a - k_b)^2 (k_b^2 - k_a^2) - (k_a + k_b)^2 (k_b^2 + k_a^2)\right] \approx$$
$$\approx \frac{\pi}{2} \cdot \frac{m \cdot k_c^2}{k_a^2 k_b^2} (k_a + k_b)^2 (k_b^2 + k_a^2) \tag{51}$$

Thus, on substituting Eqs. (17 – 18) into (45, 50 and 51), we obtain the full-width, $W(=2\Delta m)$, of the plateau as a function of $m$ and $\rho$

$$W(m, \rho) \approx \frac{8(1-\rho)(m^2 - 1)(m^2 - \rho)}{\pi \cdot m \left(\sqrt{m^2 - 1} + \sqrt{m^2 - \rho}\right)^2 \left[(m^2 - 1) + (m^2 - \rho)\right]} \quad \text{(quarter-wave layers)} \tag{52}$$

Some numerical evaluations of the function $W(m, \rho)$ are presented in Fig. 7 and Fig. 8.

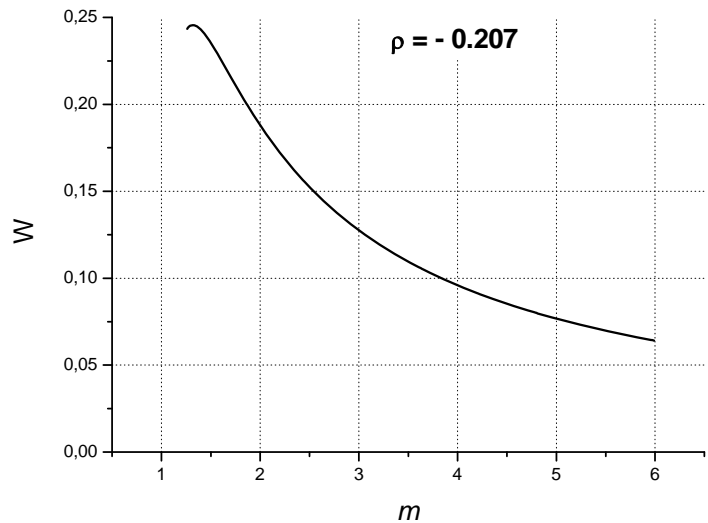

Fig. 7. Full-width of the plateau as a function of $m$. Calculations have been done for a Ni/Ti quarter-wave multilayer monochromator.

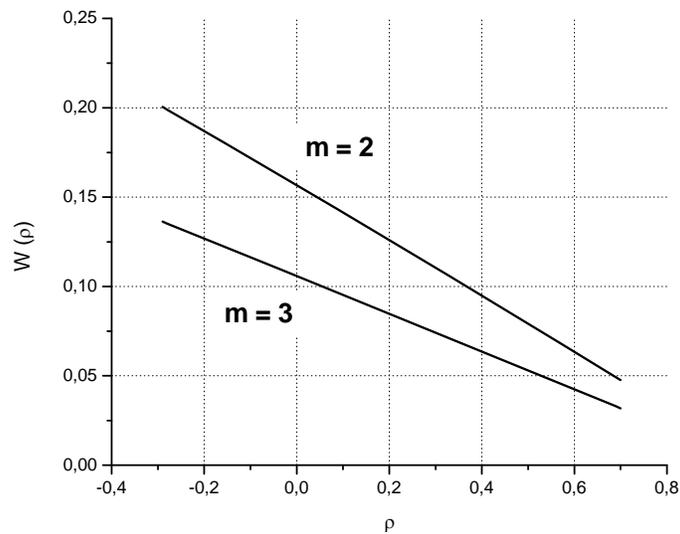

Fig. 8. Full-width of the plateau as a function of $\rho$ calculated for two different values of $m$.

One can see that the plateau is getting narrower as $m$ or/and $\rho$ increases. One could expect a multilayer with high $m$ is required to build a narrow-band monochromator. However, it is only true if a quarter-wave multilayer is considered. Looking at Fig. 3, we see that the reflectivity curve may also reach its highest value $R = 1$ for those parameters $a$ and $b$ that are well apart from the quarter-wave conditions. One may expect in that case the width of the reflectivity plateau to differ notably from the one realized by means of a quarter-wave monochromator. Table 2 presents illustratively the widths of the plateaus computed for Ni/Ti multilayer systems with different layer thicknesses *a* and *b*. All points (*a,b*) are chosen to lie in the first area with $R=1$ in Fig. 3 and fulfill the condition $Q = 0$ with $m = 3$. The first pair in Table 2 represents the quarter-wave bilayer with $k_a a = \pi/2$ and $k_b b = \pi/2$. By analogy with Fig. 6, matrix calculations have been done to simulate the reflectivity curve for each pair (*a,b*) and the width of the plateau, *W*, has been extracted from that curve. The simulations have been carried out for multilayers with a finite but sufficiently large number of layers.

Table 2
Computed parameters of multilayers with different layer thicknesses.

| a [Å] | b [Å] | W | $|C|^2$ | K |
| --- | --- | --- | --- | --- |
| 51.07 | 47.61 | 0.1277 | 0.869 | 33 |
| 22.0 | 74.7 | 0.0834 | 0.916 | 53 |
| 10.0 | 85.9 | 0.0414 | 0.958 | 108 |
| 5.0 | 90.5 | 0.0212 | 0.979 | 217 |

It is seen that the width of the plateau decreases significantly when one of the layer is getting thinner while the other layer is getting thicker. For instance, the multilayer with *a*=5Å and *b*=90.5Å features the reflectivity plateau 6 times narrower than the one of the quarter-wave multilayer. This result is of practical importance since it shows the feasibility of building a very narrow band monochromator based merely on a single reflecting multilayer. The design of such a monochromator is discussed below.

We calculate first (see Eq. (35)) the value of $|C|^2$ for each pair (*a,b*) in Table 2 and present results in the fourth column. One can see that $|C|^2$ increases as the parameter *a* decreases. This means that for a smaller *a* the penetration of the neutron wave into a multilayer becomes deeper and consequently more layers are needed to achieve a high reflectance. Knowing $|C|^2$, the number of bilayers, *K*, necessary for construction of a multilayer with the desired reflectance $R_K = R$ can be estimated directly (see, e.g., [14]). Indeed, since the intensity transmission factor of each bilayer is $|C|^2$ then the intensity transmission of *K* bilayers can be written as

$$T_K = \left(|C|^2\right)^K \qquad (53)$$

Notice, that in Eq. (53) we use $|C|^2$ derived for the case of a multilayer with an infinite number of layers. However, if the number of layers is limited but large enough to ensure a very low transmission, then the wave-function distribution within such a multilayer will be very close to the distribution obtained for the semi-infinite multilayer with the same layer parameters (see, e.g., Fig. 5). Therefore, in the case $T_K \to 0$ we may implement $|C|^2$ calculated with the use of Eqs. (35) and (40). If we neglect absorption and scattering, then holds

$$T_K = 1 - R_K \tag{54}$$

and we arrive finally at

$$K = \frac{\ln(1-R)}{\ln(|C|^2)} \tag{55}$$

The number of bilayers, $K$, required to build a multilayer monochromator with the reflectivity $R = 0.99$ at $m = 3$ is presented in the last column of Table 2. One can see that $K$ is practically inversely proportional to the width of the plateau. Thus, to build a narrow band monochromator with a reflectivity peak that is, say, 6 times narrower than the peak of the quarter-wave multilayer, one would need to build a multilayer with 6 times more bilayers. Of course, this is true only on the assumption that the layers can be of any thickness and with sharp interfaces. In practice, however, the performance of the multilayer monochromator essentially depends on the actual crystal structure of the layer materials (see comment to Table 1) as well as interfacial roughness and interdiffusion. Nevertheless, we believe that our results can serve as a guide when choosing the optimal design of a multilayer monochromator.

We turn now our attention to a quite opposite problem, namely, to the problem of design of a mirror with a very wide reflectivity band. Such mirrors are referred to as supermirrors in neutron instrumentation.

## 4. Neutron supermirror

In general, a supermirror is designed as an ordered sequence of layers of different thicknesses. In this sequence the thickness of each layer is defined by an ordering number in such a way as to ensure a high reflectivity over a broad range of the parameter $m$. We apply here the knowledge acquired in the previous sections to construct a high performance supermirror.

Following a common practice, we shall consider the deposition process where the layers are grown in such a way that the $m$-value of layers decreases from the bottom (glass surface) to the top (air surface). The layer thicknesses accordingly increase from the bottom to the top. On this basis we propose here a deposition algorithm that can be described by the following differential equation

$$\frac{d\chi}{dn} = \frac{\chi}{K^2} \tag{56}$$

where $\chi$ is the thickness of the bilayer composed of two quarter-wave layers designed for maximum reflectivity of an incident neutron beam with the fixed parameter $m$ and $K$ is the number of such bilayers necessary to attain the desired reflectivity $R$ (see Eq. (55)) for that $m$. The derivative on the left-hand side shows the variation of the thickness of the bilayer with the ordering number of bilayer $n$. Eq. (56) was developed intuitively on the basis of a general idea that came from the analysis of Eqs. (25) and (52). Indeed, from Eq. (25) it follows that the layer thickness may vary slightly without affecting the reflectivity of a multilayer at a given $m$. In our case, according to Eq. (56), the thickness of the bilayer $\chi_K$ with the ordering number $n = K$ differs from the thickness $\chi_1$ by the amount of

$$\chi_1 - \chi_K \approx \frac{\chi_1}{K} \tag{57}$$

which is not far from the estimate obtained with Eq. (25). Thus, one may expect that all $K$ bilayers contribute to the maximum reflectivity at the given $m$. On the other hand, we have seen that the maximum reflectivity represents a plateau over some interval around fixed $m$ (see Eq. (52)). Therefore, we tried to construct Eq. (56) so that it can ensure both the maximum reflectivity within such $m$-intervals and smooth matching between subsequent $m$-intervals. Let us turn now to the solution of this equation.

After insertion of Eq. (55) into (56) we get

$$\frac{d\chi}{dn} = \chi \frac{\ln^2(|C|^2)}{\ln^2(1-R)}. \tag{58}$$

With the use of Eqs. (17-18) we can write the thickness of the quarter-wave bilayer $\chi$ as

$$\chi = a + b = \frac{\pi}{2k_a(m)} + \frac{\pi}{2k_b(m)} = \frac{\pi}{2k_c}\left(\frac{1}{\sqrt{m^2-1}} + \frac{1}{\sqrt{m^2-\rho}}\right). \tag{59}$$

Then, from Eq. (40) we have

$$m = \sqrt{\frac{1-|C|^2\rho}{1-|C|^2}} \tag{60}$$

and, after substitution this into Eq. (59), we obtain

$$\chi = \frac{\pi}{2k_c} \cdot \frac{1+|C|}{|C|} \cdot \sqrt{\frac{1-|C|^2}{1-\rho}} \tag{61}$$

Finally, Eq. (58) reduces to

$$\frac{d|C|}{dn} = \frac{|C|(1-|C|^2)}{|C|-|C|^2-1} \cdot \frac{\ln^2(|C|^2)}{\ln^2(1-R)} \tag{62}$$

The thickness of the bilayer as a function of the ordering number, $\chi = \chi(n)$, can now be estimated by solving simultaneously Eqs. (58) and (62). The thickness of each layer, $a$ and $b$, in the quarter-wave bilayer can be evaluated from the set of equations

$$\begin{cases} a + b = \chi \\ \dfrac{k_a a}{k_b b} = |C|\dfrac{a}{b} = 1 \end{cases} \quad (63)$$

Thus we get

$$\begin{cases} a = \dfrac{1}{1+|C|}\chi \\ b = \dfrac{|C|}{1+|C|}\chi \end{cases} \quad (64)$$

Since a multilayer is a system with discrete layers, it is necessary to rewrite our equations in a discrete form. Finally, the proposed algorithm for design of a supermirror can be presented as an iteration process described by the following set of equations:

$$|C|_{n+1} = |C|_n + \frac{|C|_n\left(1-|C|_n^2\right)}{|C|_n - |C|_n^2 - 1} \cdot \frac{\ln^2\left(|C|_n^2\right)}{\ln^2(1-R)} \quad (65a)$$

$$\chi_{n+1} = \chi_n + \chi_n \cdot \frac{\ln^2\left(|C|_n^2\right)}{\ln^2(1-R)} \quad (65b)$$

$$a_n = \frac{1}{1+|C|_n}\chi_n \quad (65c)$$

$$b_n = \frac{|C|_n}{1+|C|_n}\chi_n \quad (65d)$$

In Eq. (65) the subscript index $n$ denotes the ordering number.
The input parameters are: the maximum value $m_{max}$ that represents the upper limit of the $m$-range of high reflectivity; the material parameters $k_c$ and $\rho$ (see Eqs. (12) and (19)) and the desired value of the highest reflectivity $R$ which is taken to be the same for all $m$ within the entire $m$-range. The initial values of $|C|$ and $\chi$ can be obtained directly from Eqs. (40) and (59) with $m = m_{max}$. The iteration process should be interrupted when $|C|_n$ approaches its smallest positive value.

    For the purposes of illustration we have implemented this algorithm for design of a Ni/Ti supermirror with $m_{max} = 3$ deposited on a Si wafer. As a preliminary, the supermirror is assumed to be constructed of layers with no absorption or scattering. The parameter R was chosen to be 0.99. In this case the algorithm (65) ended up with 276 bilayers (552 layers in total). The total thickness of the multilayer comprises 3.7 μm. It is worth noting that the top layers can be subject to manual correction. It happens if the calculated thicknesses of these layers are too big to be practical. To make the design practical, one can manually remove those layers from the designed structure and put on the very top (air surface) a Ni-layer of about 600 Å thickness. Fig. 9 shows

the reflectivity spectrum computed for the designed supermirror. The calculations have been done with the use of the matrix method.

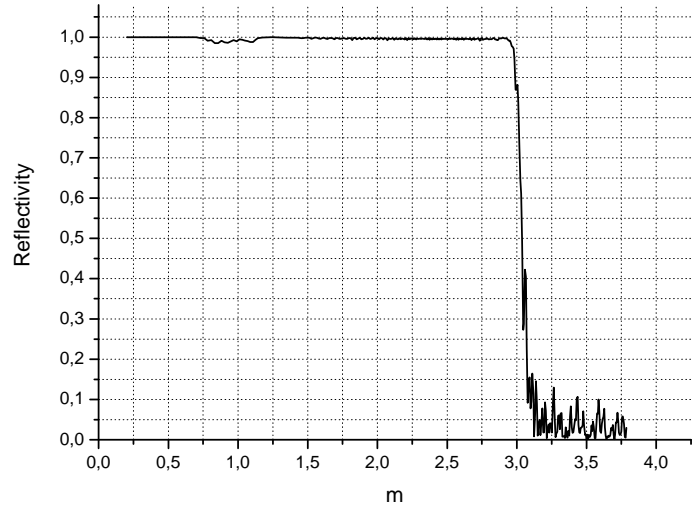

Fig. 9. Calculated reflectivity of the Ni/Ti supermirror with $m_{max} = 3$ and $R = 0.99$. Absorption and scattering in the layers are neglected.

One can see that the reflectivity curve in Fig. 9 fits very well the specified parameters of the supermirror.
    As the next step towards practical design one has to take into account absorption and scattering of neutrons in the layers. This can be done by substituting a complex value for the neutron potential $U = U' - iU''$ in Eqs. (2,3). The imaginary part of the potential can be written as ([21, 22]):

$$U''(k) = \frac{\hbar^2 k}{2m_n} N\sigma_{tot}(k) \qquad (66)$$

Here $\sigma_{tot}$ is the total cross section that describes the attenuation of the neutron wave within a bulk medium and $k$ is the magnitude of the total neutron wave vector. One can then redefine the scattering length density in Eqs. (1 - 3) as follows:

$$(Nb) = (Nb)' - i(Nb)'' \qquad (67)$$

with the imaginary part

$$(Nb)'' = N\frac{k}{4\pi}\sigma_{tot}(k) \qquad (68)$$

It should be noted that the imaginary part depends on the magnitude of the total wave vector $k$ rather than the normal component. Therefore, one may expect that reflectivity curves measured for the same supermirror but at different instruments may differ from each other depending on the wavelength of an incident neutron beam at the instrument. An additional point to emphasize is that the reflectivity may vary significantly in the vicinity of the Bragg reflections from crystalline materials comprising the multilayer. For the Ni/Ti supermirror this effect should be most pronounced near the Bragg cut-off for nickel ($\lambda = 4$Å [23]). One can expect that it will manifest itself as some dip at 4Å in the continuous spectrum of a neutron beam passed through a long Ni/Ti neutron guide.

It might be well to point out that the value of the total cross section in Eqs. (66) and (68) is well defined only for a bulk medium where all crystalline grains are randomly oriented. In the case of thin metallic films deposited on each other a preferential orientation of the grains may take place and this would result in some variation of the total cross section [24].

To be specific, in the subsequent calculations we assume the incident neutron wavelength to be 1.8Å and there is no preferential orientation of the grains. With the use of tabulated data for neutron cross sections [25] we obtain

$$(Nb)_{Ni} = 9.41 \cdot 10^{-6} - i5.81 \cdot 10^{-9} \ (\text{Å}^{-2}) \qquad (\lambda = 1.8\text{Å})$$
$$(Nb)_{Ti} = -1.95 \cdot 10^{-6} - i1.63 \cdot 10^{-9} \ (\text{Å}^{-2}) \qquad (\lambda = 1.8\text{Å})$$

Fig. 10 displays the reflectivity curve calculated for the supermirror with the same structure as in Fig. 9 but with the imaginary part of the potential taken into account. Here, the deviation of the reflectivity from the unity is clearly seen.

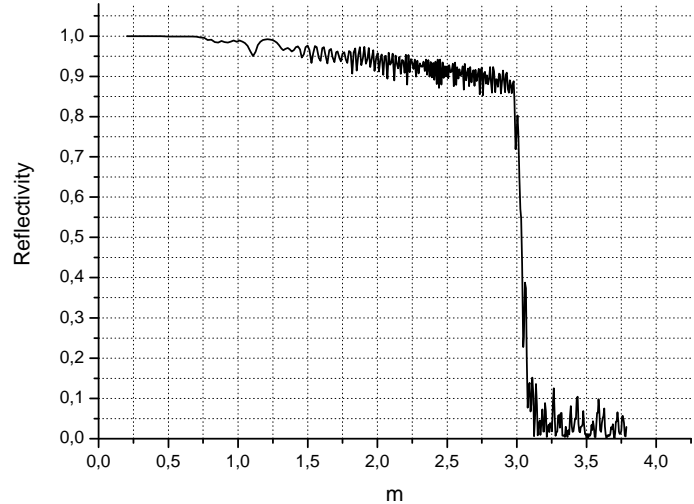

Fig. 10. Calculated reflectivity of the Ni/Ti supermirror with $m_{max} = 3$ and $R = 0.99$. Attenuation in the layers is taken into consideration.

Thus, the wave attenuation in the layers sets an upper limit upon the reflectivity of a supermirror. Since this restriction is unavoidable it seems reasonable to relax the requirement for the design of a supermirror and reduce the parameter $R$ in Eq. (65) in such a way that the total reflectivity curve will not differ noticeably from the curve in Fig. 10 (the same procedure was implemented in [17]). As an example, Fig. 11 shows the reflectivity curve calculated for the supermirror with the attenuation in the layers and with the geometry based on the algorithm (65) but with the reduced parameter $R = 0.98$. This small variation of the parameter $R$ resulted in significant decrease in the necessary number of bilayers down to 201 (402 layers in total), that is roughly 27% less against the supermirror designed with $R = 0.99$ (see above). The total thickness of the multilayer in this case comprises 2.7 μm.

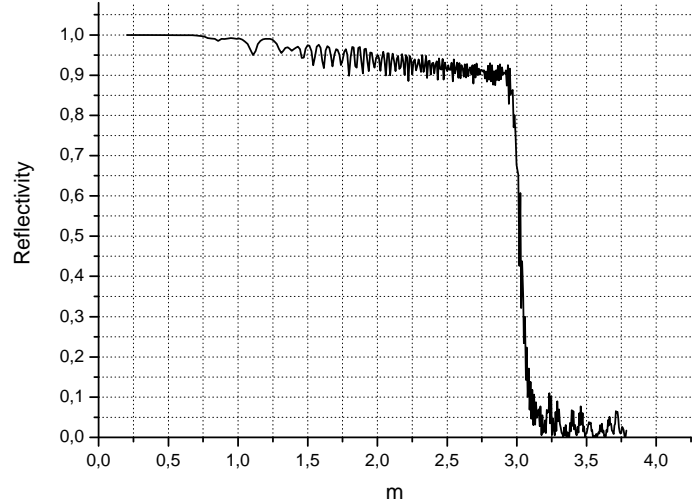

Fig. 11. Calculated reflectivity of the Ni/Ti supermirror with $m_{max} = 3$ and $R = 0.98$. Attenuation in the layers is taken into consideration.

The reflectivity curve calculated under these conditions is not worse than the one in Fig. 10. Moreover, the edge reflectivity (i.e., in the vicinity of m = 3) in Fig. 11 became even higher than it was in Fig. 10 with $R = 0.99$. Such enhancement can be explained by the fact that the sinking of the reflectivity curve due to decrease of the parameter $R$ in Eq. (65) is overcompensated by the reduction of the number of layers and hence the total attenuation of the neutron waves in the layers.

     It seems logical to try further simplifying the design. Indeed, so far we have only considered multilayers with sharp interfaces and the attenuation within the layers was the only source of theoretical restriction on the reflectivity. However, in practice the production of high-reflectivity supermirrors is affected by the roughness of the interfaces as well (see, e.g., [18, 20, 26]). So, one can expect that the reflectivity measured with a real supermirror will differ from the theoretical value where only the total cross section related to a bulk medium was taken into account. Therefore, it might be thought that the further reduction of the parameter $R$ would be acceptable and could result in even simpler design of the supermirror. Fig. 12 shows the reflectivity curve computed for the supermirror designed according to the algorithm (65) with $R = 0.97$. The multilayer is now composed of only 164 bilayers (328 layers in total).

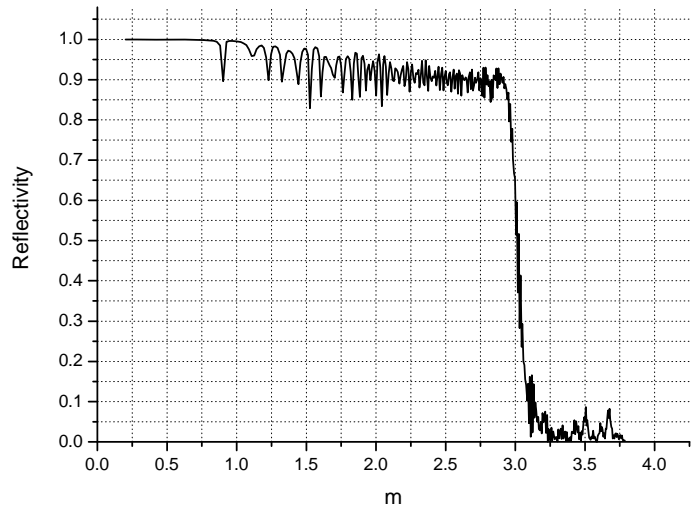

Fig. 12. Calculated reflectivity of the Ni/Ti supermirror with $m_{max} = 3$ and $R = 0.97$. Attenuation in the layers (total cross section) is taken into consideration.

We see that the edge reflectivity in Fig. 12 is still about 90%, although the curve is not as smooth as in Fig. 11 and Fig. 10. We ascribe this to the fact that the multilayer structure designed according to Eq. (65) with $R = 0.97$ is getting coarser with the noticeable mismatch between the sequential layers. One can hope that even in this case some enhancement of the reflectivity curve can be achieved by setting the parameter $R$ not as a constant but as a function $R(m)$. The evaluation of the appropriate function $R(m)$ can be the subject of further study.